# Weight-conserving characterization of complex functional brain networks

Mikail Rubinov[1,2,3] and Olaf Sporns[4]

*Abstract:* **Complex functional brain networks are large networks of brain regions and functional brain connections. Statistical characterizations of these networks aim to quantify global and local properties of brain activity with a small number of network measures. Important functional network measures include measures of modularity (measures of the goodness with which a network is optimally partitioned into functional subgroups) and measures of centrality (measures of the functional influence of individual brain regions). Characterizations of functional networks are increasing in popularity, but are associated with several important methodological problems. These problems include the inability to characterize densely connected and weighted functional networks, the neglect of degenerate topologically distinct high-modularity partitions of these networks, and the absence of a network null model for testing hypotheses of association between observed nontrivial network properties and simple weighted connectivity properties. In this study we describe a set of methods to overcome these problems. Specifically, we generalize measures of modularity and centrality to fully connected and weighted complex networks, describe the detection of degenerate high-modularity partitions of these networks, and introduce a weighted-connectivity null model of these networks. We illustrate our methods by demonstrating degenerate high-modularity partitions and strong correlations between two complementary measures of centrality in resting-state functional magnetic resonance imaging (MRI) networks from the 1000 Functional Connectomes Project, an open-access repository of resting-state functional MRI datasets. Our methods may allow more sound and reliable characterizations and comparisons of functional brain networks across conditions and subjects.**

## Introduction

Large-scale functional brain networks are networks of brain regions and functional connections – coactivations or correlations – between pairs of these regions. Complex functional brain networks are large and extensive networks of nontrivially interacting brain regions that often serve as maps of global brain activity (Bullmore and Sporns, 2009). Interactions between regions in complex functional networks vary in magnitude from large to small, and vary in sign from positive to negative. In contrast, the more traditional "simple" functional networks are smaller groupings of strongly and mutually correlated regions that often serve as maps of specialized functional systems (Fox and Raichle, 2007). Simple functional networks form highly connected modules – components or subnetworks – within complex functional networks (Meunier et al., 2010).

Statistical characterizations of complex functional networks attempt to quantify global and local properties of these networks with a small number of network measures (Stam and Rejneveld, 2007; Bullmore and Sporns, 2009; Rubinov and Sporns, 2010). Important functional network measures include measures of modularity (measures of the goodness with which a network is optimally partitioned into modules) and measures of centrality (measures of the functional influence of individual brain regions). For instance, recent characterizations partitioned complex resting-state functional MRI networks into modules which correspond to visual, attention, default mode and other "simple" networks (Fair et al., 2008; He et al., 2009; Meunier et al., 2009) and identified prominent central brain regions in heteromodal association areas (Achard et al., 2006; Buckner et al., 2009). Interestingly, several studies reported alterations of complex functional network topology in neurological and psychiatric disorders, such as Alzheimer's disease (Supekar et al., 2008) and schizophrenia (Lynall et al., 2010).

Despite these promising findings, current characterizations of complex functional networks are associated with several methodological problems. Firstly, most current network measures are optimally suited for sparse and binary networks and are less well suited for dense and weighted networks. This often necessitates the conversion of dense and weighted complex functional networks to sparse and binary form. Such conversions are made by defining a weight threshold, setting all suprathreshold connection weights to 1 and all subthreshold weights to 0. These thresholding and binarizing manipulations are associated with loss of information and are often arbitrarily made. Secondly, current modularity studies typically describe only one high-modularity partition. Recently, Good et al. (2010)

1. Black Dog Institute and School of Psychiatry, University of New South Wales, Sydney, Australia; 2. Mental Health Research Division, Queensland Institute of Medical Research, Brisbane, Australia; 3. CSIRO Information and Communication Technologies Centre, Sydney, Australia; 4. Department of Psychological and Brain Sciences, Indiana University, Bloomington, USA.

Corresponding author: Mikail Rubinov, Mental health research division, Queensland institute of medical research, PO Royal Brisbane Hospital, QLD 4029, Australia. Email: m.rubinov@student.unsw.edu.au

analytically showed that modular complex networks are likely to have many topologically distinct, high-modularity partitions. The presence of such "degenerate" partitions makes it less meaningful to focus on only one potentially unrepresentative case. Degenerate partitions have not been previously reported in functional brain networks, although similar concepts have been considered in the context of cluster stability (Bellec et al., 2010). Thirdly, no satisfactory null model of complex functional networks is currently available. A null model of functional networks should allow to test null hypotheses of association between observed nontrivial network properties, such as high modularity and degeneracy, and simple network properties, such as basic organization of weighted connectivity.

In this study, we describe a set of methods that aim to overcome the above methodological problems. Firstly, we generalize measures of modularity and centrality to fully connected networks with positive and negative weights. Secondly, we describe the detection of degenerate high-modularity partitions of these networks. Thirdly, we introduce a weighted null model of these networks. We illustrate our methods by characterizing resting-state functional MRI networks from the 1000 Functional Connectomes Project, an open-access repository of resting-state functional MRI datasets (Biswal et al., 2010).

### Methods

In this section we provide mathematical definitions of our proposed measures and algorithms. The Brain Connectivity Toolbox (http://www.brain-connectivity-toolbox.net), an open-access Matlab network analysis toolbox which we maintain, contains software to compute these measures. We encourage readers unfamiliar with complex networks methodology to consult our less technical and more accessible recent overview (Rubinov and Sporns, 2010).

Complex functional brain networks are fully connected, undirected, positively and negatively weighted networks of $n$ nodes and $\frac{1}{2}n(n-1)$ connections. We denote the presence of a positively weighted connection between nodes $i$ and $j$ in these networks with $a_{ij}^+ = 1$ and $a_{ij}^- = 0$, and we denote the weight of this connection with $w_{ij}^+ \in (0,1]$, $w_{ij}^- = 0$. Equivalently, we denote the presence of a negatively weighted connection between nodes $i$ and $j$ with $a_{ij}^- = 1$, $a_{ij}^+ = 0$, and the weight of this connection with $w_{ij}^- \in (0,1]$, $w_{ij}^+ = 0$. The degree of node $i$, $k_i^{\pm} = \sum_j a_{ij}^{\pm}$, is the number of positive or negative connections of $i$. The strength of node $i$, $s_i^{\pm} = \sum_j w_{ij}^{\pm}$, is the sum of positive or negative connection weights of $i$. The total weight, $v^{\pm} = \sum_{ij} w_{ij}^{\pm}$, is the sum of all positive or negative connection weights (counted twice for each connection).

*Measures of modularity in networks with positive and negative weights*

A modularity partition is the complete subdivision of the network into nonoverlapping modules (Fortunato, 2010). Measures of modularity quantify the goodness of modularity partitions (Newman, 2004). In networks with no negative weights, a de-facto-standard measure of modularity is the average difference between present within-module connection weights $w_{ij}^+$ and chance-expected within-module connection weights $e_{ij}^+$,

$$Q^+ = \frac{1}{v^+} \sum_{ij} (w_{ij}^+ - e_{ij}^+) \delta_{M_i M_j},$$

where $e_{ij}^{\pm} = \frac{s_i^{\pm} s_j^{\pm}}{v^{\pm}}$, $\delta_{M_i M_j} = 1$ when $i$ and $j$ are in the same module and $\delta_{M_i M_j} = 0$ otherwise (Newman, 2006). Partitions with high $Q^+$ therefore have larger than chance-expected total positive within-module weight. The factor $\frac{1}{v^+}$ rescales the maximized $Q^+$ to the range of [0,1].

In this study we assume that both positively and negatively weighted connections provide useful information about the goodness of modularity partitions. Specifically, we consider positively weighted connections to represent similar activation patterns and hence to support placement of positively connected pairs of nodes in the same module. The measure $Q^+$ reflects this contribution of positive weights. On the other hand, we consider negatively weighted connections to represent distinct activation patterns or antiphase coupling and hence to support placement of negatively connected pairs of nodes in distinct modules. An analogous measure $Q^-$ reflects this contribution of negative weights,

$$Q^- = -\frac{1}{v^-} \sum_{ij} (w_{ij}^- - e_{ij}^-) \delta_{M_i M_j}.$$

We now consider a simple modularity measure for networks with positive and negative weights,

$$Q^{\text{simple}} = Q^+ + Q^-.$$

This measure is problematic because it does not rescale $Q^+$ and $Q^-$. For instance, the presence of precisely one negative connection in the network always inappropriately contributes 0.5 to $Q^{\text{simple}}$, irrespectively of other connectivity (this follows from the definition of $Q^-$). A number of recently proposed modularity measures attempt to overcome this problem by rescaling $Q^+$ and $Q^-$ by the total connection weight $v^+ + v^-$. For instance, Traag and Bruggeman (2009) generalize the standard modularity measure as

$$Q^{\text{TB}} = \frac{1}{v^+ + v^-} \sum_{ij} [(w_{ij}^+ - \gamma^+ e_{ij}^+) - (w_{ij}^- - \gamma^- e_{ij}^-)] \delta_{M_i M_j},$$

where $\gamma^+$ and $\gamma^-$ are module size parameters. Gomez, Jensen and Arenas (2009) consider a special case of $Q^{\text{TB}}$ for $\gamma^{\pm} = 1$,

$$Q^{\text{GJA}} = \frac{1}{v^+ + v^-} \sum_{ij} [(w_{ij}^+ - e_{ij}^+) - (w_{ij}^- - e_{ij}^-)] \delta_{M_i M_j}.$$

Finally, Kaplan and Forrest (2008) generalize the standard modularity measure slightly differently, by additionally rescaling the chance-expected within-module connection weights,

$$Q^{\text{KF}} = \frac{1}{v^+ + v^-} \sum_{ij} [(w_{ij}^+ - e'^+_{ij}) - (w_{ij}^- - e'^-_{ij})] \delta_{M_i M_j},$$

where $e'^{\pm}_{ij} = \frac{s_i^{\pm} s_j^{\pm}}{v^+ + v^-}$ replaces $e^{\pm}_{ij} = \frac{s_i^{\pm} s_j^{\pm}}{v^{\pm}}$. All these modularity measures reduce to the standard measure when there are no negative weights in the network.

*An asymmetric measure of modularity in networks with positive and negative weights*

All the above generalizations treat $Q^+$ and $Q^-$ symmetrically and are hence based on the assumption that positive and negative weights are equally important. Here we argue that this assumption is neurobiologically problematic, because the role and importance of positive and negative weights in functional networks is intrinsically unequal. Positive weights associate nodes with modules explicitly, while negative weights associate nodes with modules implicitly, by dissociating nodes from other modules. Empirically, high-$Q^-$ modularity partitions are objectively less optimal than high-$Q^+$ modularity partitions (see Results).

The unequal importance of positive and negative weights in modularity-partition determination motivates an asymmetric generalization of the standard measure. We hence define

$$\begin{aligned} Q^* &= Q^+ + \frac{v^-}{v^+ + v^-} Q^- \\ &= \frac{1}{v^+} \sum_{ij} (w_{ij}^+ - e_{ij}^+) \delta_{M_i M_j} \\ &\quad - \frac{1}{v^+ + v^-} \sum_{ij} (w_{ij}^- - e_{ij}^-) \delta_{M_i M_j}. \end{aligned}$$

Our definition explicitly makes the contribution of negative weights auxiliary to the contribution of positive weights. For instance, increase in positive weights reduces the influence of negative weights in $Q^*$, unlike in $Q^{\text{simple}}$. On the other hand, increase in negative weights does not reduce the influence of positive weights in $Q^*$, unlike in $Q^{\text{TB}}$ (hence $Q^{\text{GJA}}$) and $Q^{\text{KF}}$. In networks with approximately equal numbers of positive and negative weights, the influence of positive weights in $Q^*$ will be twice as large as the influence of negative weights. High-$Q^*$ partitions should theoretically have most positive weights within modules, and most negative weights between modules. Maximized values of $Q^*$ are in the range of [0,1].

We now illustrate the advantage of our measure with an example. We consider a network of 100 nodes and four equally sized modules. The network has a maximal number of positive weights within modules, and no positive weights between modules. Initially, the network has no negative weights. For simplicity we set all $w_{ij}^{\pm} = 1$. We now randomly add negative weights to this network until the network becomes fully connected (Figure 1a), and observe the resulting changes in values of the modularity measures (Figures 1b-e). We note that only $Q^*$ behaves as we intuitively expect: the contribution of positive weights does not change, while the contribution of negative weights gradually increases. The contribution of positive weights to $Q^{\text{simple}}$ is paradoxically large when there are few negative weights, while the contribution of positive weights to $Q^{\text{TB}}$ (hence $Q^{\text{GJA}}$) and $Q^{\text{KF}}$ paradoxically decreases, even as the number and organization of positive weights does not change.

*Evaluation of the goodness of modularity partitions*

The goodness of modularity partitions may in general be quantified with many measures. In addition to $Q^+$ and $Q^-$, these measures include the proportion of within-module positive and negative weights,

$$F^{\pm} = \pm \frac{1}{v^{\pm}} \sum_{ij} w_{ij}^{\pm} \delta_{M_i M_j},$$

and the geometric mean of balanced and unbalanced within-module triangle weights (Onnela et al., 2005),

$$T^{\pm} = \pm \frac{\sum_{h,i,j} (w_{hi}^+ w_{hj}^+ w_{ij}^{\pm})^{\frac{1}{3}} \delta_{M_h M_i M_j}}{\sum_{h,i,j} (a_{hi}^+ a_{hj}^+ a_{ij}^{\pm}) \delta_{M_h M_i M_j}}.$$

where $\delta_{M_h M_i M_j} = \delta_{M_h M_i} \delta_{M_i M_j}$. $T^{\pm}$ is a generalization of the weighted clustering coefficient to networks with positive and negative weights and is related to the within-module density. Good modularity partitions should have an unexpectedly large proportion of clustered, positively weighted connections within modules (high $Q^+$, $F^+$ and $T^+$, respectively), and an unexpectedly small proportion of unclustered, negatively weighted connections within modules (high $Q^-$, $F^-$ and $T^-$, respectively). Note that $Q^+$, $Q^-$, $F^+$ and $T^+$ lie in the range of [0,1], while $F^-$ and $T^-$ lie in the range of [−1,0].

*Detection of degenerate high-modularity partitions*

Degenerate high-modularity partitions are topologically distinct modularity partitions with similarly high values of the modularity (Good et al., 2010). We argue that characterization of the modularity in functional networks should comprehensively describe the set of these partitions. In this study, we searched for degenerate partitions in two steps:

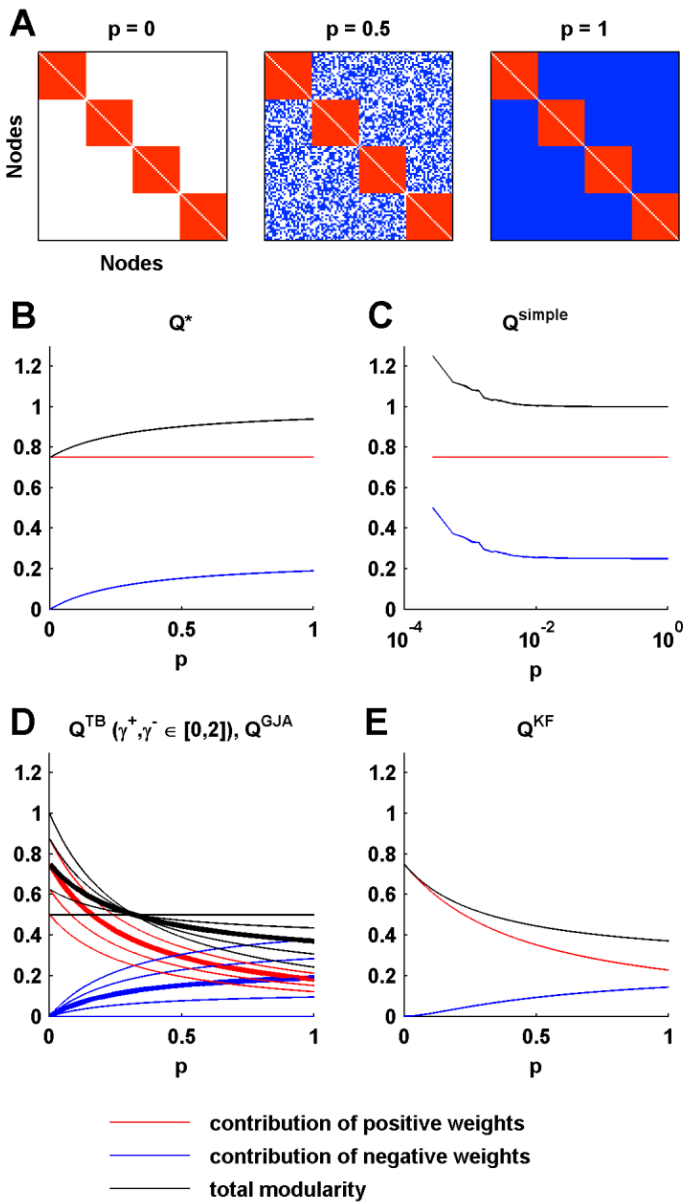

**Figure 1.** Conceptual advantages of an asymmetric modularity measure. (a) A network with a maximal number of within-module positively weighted (red) connections and an increasing number (from left to right) of between-module negatively weighted (blue) connections. (b)-(e) Contributions to modularity of positive (red) and negative (blue) weights, and the total modularity (black) as a function of the proportion of negative between-module connections. (b) $Q^*$, (c) $Q^{simple}$ (note the semilogarithmic scale), (d) $Q^{TB}$ for $\gamma^\pm \in [0,2]$; results for $Q^{GJA}$ ($\gamma^\pm = 1$) are shown in bold, (e) $Q^{KF}$. Values are averages of 10 simulations.

1. We first identified several seed partitions for subsequent systematic exploration of degeneracy. We identified these partitions with a popular greedy modularity-maximization algorithm (Blondel et al., 2008) and a fine-tuning algorithm (Sun et al., 2009), but in principle many other modularity-maximization algorithms may be used (Fortunato, 2010). Most modularity-maximization algorithms search for high-modularity partitions heuristically and discover slightly different partitions from run to run. Accordingly, the Blondel et al.

(2008) algorithm estimates optimal module affiliation for each node by sequentially and repeatedly examining nodes in random order. This randomness is associated with potential variations in discovered high-modularity partitions. We applied this algorithm 1000 times to each examined network to obtain 1000 preliminary seed partitions. We further refined these partitions with a Kernighan-Lin (1970)-based fine-tuning algorithm (Sun et al., 2009). More specifically, we iteratively refined each partition by computing changes in the modularity associated with all potential between-module node moves, including moves that resulted in creation of new modules. We then considered a node associated with an optimal move, and made the corresponding move if it increased the modularity. At each iteration of the algorithm we considered each node exactly once, and we repeatedly iterated until no further moves increased the modularity. These refined partitions constituted our seed partitions for step 2 and represent peaks in the modularity landscape of the network. Seed partitions were not all necessarily distinct.

2. We next systematically searched for degenerate high-modularity partitions by rerunning one further iteration of the fine-tuning algorithm on seed partitions, and incorporating into this iteration random moves of randomly chosen (but not previously considered) nodes. We defined partitions to be degenerate when the modularity of these partitions was in the top 1% of the estimated maximum. We examined 1 million candidate degenerate partitions by varying the probability of random moves for each seed partition from 0 (no randomization) to $5 \times 10^{-2}$ in $5 \times 10^{-5}$ increments (we empirically determined that the lower bound captured an overwhelming majority of degenerate partitions in the studied networks).

We quantified the similarity of individual partitions with the variation of information, a popular information-theoretic measure of distance in partition space (Meila, 2007). To compute the variation of information, we first define the entropy associated with a partition $M$ as

$$H(M) = -\sum_{u \in M} P(u) \log P(u),$$

where the sum is over all modules in $M$, $P(u) = \frac{n_u}{n}$ and $n_u$ is the number of nodes in module $u$. We then analogously define the mutual information between two partitions $M$ and $M'$ as

$$I(M, M') = \sum_{u \in M} \sum_{u' \in M'} P(u, u') \log \frac{P(u, u')}{P(u)P(u')},$$

where $P(u, u') = \frac{n_{uu'}}{n}$ and $n_{uu'}$ is the number of nodes that are simultaneously in module $u$ of partition $M$, and in module $u'$ of partition $M'$. We finally define the variation of information as

$$VI = \frac{1}{\log n}[H(M) + H(M') - 2I(M,M')],$$

where the factor $\frac{1}{\log n}$ rescales the variation of information to the range of [0,1], such that $VI = 0$ corresponds to equal partitions, and $VI = 1$ corresponds to maximally distant partitions (Karrer et al., 2008).

We note two things about our algorithm. Firstly, the 1% threshold is conservative and may potentially miss meaningful partitions with lower modularity. Secondly, the study of modularity degeneracy is still in its infancy, and efficient and exhaustive search algorithms are only beginning to be developed (Good et al., 2010; Duggal et al., 2010). Consequently, while we find a reasonably large number of degenerate partitions with our algorithm, we cannot claim that the algorithm searches exhaustively. On the other hand, the knowledge of all possible partitions may not be needed as long as the discovered partitions are evenly sampled and are representative of the whole partition set.

We now illustrate the application of our algorithm to simple model networks. We first consider, as previously, a network of 100 nodes and four equally sized modules. The network has a maximal number of positive weights within modules, and a maximal number of negative weights between modules (Figure 2a). We randomize this network and examine the number of discovered degenerate partitions as a function of randomization. Figure 2b shows that the number of discovered partitions is close to 1 at low levels of randomization (in modular networks), is relatively low at large levels of randomization (in random networks) and peaks at the transition from modular to random network topology. Figure 2c shows that the distance between discovered partitions is low for modular networks, rapidly rises at the transition, and remains high for random networks.

To probe the evenness with which our algorithm samples degenerate partitions, we apply this algorithm to a lattice network (Figure 2d) and obtain a within-module connectivity likelihood matrix by averaging the topology of all discovered partitions for this network. Connections with high values in this matrix are hence likely to be located inside modules. Figure 2e shows that the likelihood matrix accurately reconstructs the homogeneous lattice topology, and hence suggests an even sampling of degenerate partitions of the lattice.

*Measures of centrality in networks with positive and negative weights*

Central brain regions have functional connections to many diverse regions. These properties are intuitively captured in measures of regional connection strength and regional connection diversity. We define the normalized connection strength as

$$s_i'^{\pm} = \frac{1}{n-1}s_i^{\pm},$$

and we define the normalized connection diversity as

$$h_i^{\pm} = -\frac{1}{\log m}\sum_{u \in M} p_i^{\pm}(u)\log p_i^{\pm}(u),$$

where $p_i^{\pm}(u) = \frac{s_i^{\pm}(u)}{s_i^{\pm}}$, $s_i^{\pm}(u)$ is the strength of node $i$ within module $u$ (the total weight of connections of $i$ to all nodes in $u$), and $m$ is the number of modules in modularity partition $M$. The factors $\frac{1}{n-1}$ and $\frac{1}{\log m}$ rescale $s_i'^{\pm}$ and $h_i^{\pm}$ to the range of [0,1]. Central nodes should have high $s_i'^{+}$ and $h_i^{+}$ but low $s_i'^{-}$ and $h_i^{-}$. As above, we consider the strength and diversity of positively weighted connections to be more important than the strength and diversity of negatively weighted connections. We accordingly generalize $s_i'^{+}$ and $h_i'^{+}$ as

$$s_i'^{*} = s_i'^{+} - \left(\frac{s_i^{-}}{s_i^{+}+s_i^{-}}\right)s_i'^{-}, \qquad h_i^{*} = h_i^{+} - \left(\frac{s_i^{-}}{s_i^{+}+s_i^{-}}\right)h_i^{-},$$

where we rescale the contribution of negative weights by $\frac{s_i^{-}}{s_i^{+}+s_i^{-}}$, rather than by $\frac{v^{-}}{v^{+}+v^{-}}$, due to the local nature of centrality measures. Our generalizations favor nodes with high but equal positive and negative strength and diversity over nodes with low and equal positive and negative strength and diversity. Both $s_i'^{*}$ and $h_i^{*}$ are in the range of [−1,1].

Our measures of centrality have clear parallels, but also important differences, with the commonly used "cartographic" classification of node roles in complex networks (Guimera and Amaral, 2005). The cartographic classification of node roles likewise quantifies connection strength and connection diversity, but uses a measure of the within-module strength for the former, and a measure of the participation coefficient for the latter. Here we use the total connection strength, rather than the within-module connection strength, because functional centrality usually presumes global integration, and because within-module strength profiles should be similar by transitivity of correlations. In addition, we compute connection diversity with the normalized Shannon (1948) entropy, rather than the participation coefficient, or equivalently the Simpson (1949) index. Shannon entropy and Simpson index are similar measures of diversity (e.g. Keylock, 2005) but importantly differ in the upper bound on maximally central nodes. The normalized Shannon entropy has a constant upper bound of 1, while the Simpson index has a variable upper bound of $1 - \frac{1}{m}$. The upper bound of the Simpson index may hence differ for partitions of different networks, and for different partitions of the same network. These considerations led us to adopt the Shannon entropy as a more consistent measure of diversity.

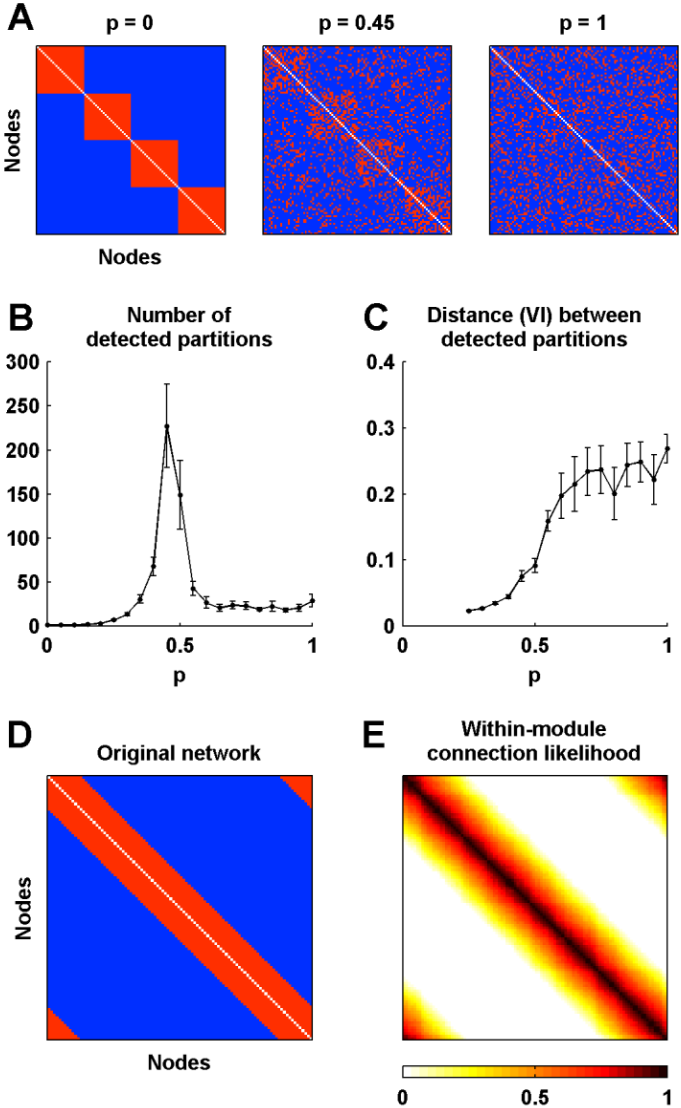

**Figure 2.** Detection of degenerate high-modularity partitions. (a) Modular, intermediate and random networks with positively weighted (red) and negatively weighted (blue) connections. (b) Number of discovered partitions and (c) average distance between discovered partitions as a function of network randomization. Error bars represent the standard error of the mean from 10 simulations. (d) A lattice network and (e) the estimated within-module connection likelihood of the lattice network, computed as the average over all discovered degenerate partitions. The number of discovered degenerate partitions for the lattice was approximately 320.

### Degree-, weight- and strength-preserving null model

Nontrivial properties of network topology, such as high modularity and degeneracy, can only be conclusively claimed through comparisons with appropriate null models. A standard null model of binary networks is a random network with preserved degrees. To our knowledge, there is no corresponding null model of weighted networks. Here we introduce such a model by generalizing the binary null model to preserve node degrees, preserve connection weights and closely approximate node strengths. The corresponding null hypothesis hence asserts that observed network properties are associated with degree, weight and strength properties of the network.

Our algorithm consists of two steps:

1. We first randomize network connections in a way that preserves positive and negative degrees. There are a number of ways to achieve this and we use a simple popular heuristic known as the connection-switching method (e.g. Wormald, 1999). We choose four nodes $i_1$, $i_2$, $i_3$, $i_4$ at random, such that $a^+_{i_1 i_2} = a^+_{i_3 i_4} = 1$ and $a^-_{i_1 i_4} = a^-_{i_3 i_2} = 1$. We then set $a^+_{i_1 i_4} = a^+_{i_3 i_2} = 1$ and $a^-_{i_1 i_2} = a^-_{i_3 i_4} = 1$. We iteratively repeat this process until the network is randomized.

2. We next associate original network weights with connections in a way that closely approximates the positive and negative strengths. This procedure is equivalent for positive and negative weights. For positive weights, we begin by ranking all original $w^+_{ij}$ by magnitude. At the same time, we associate all $a^+_{ij} = 1$ in the network with $\hat{w}^+_{ij} = 0$, and we rank all $a^+_{ij}$ by the expected weight magnitude, $\hat{e}^+_{ij} \propto (s^+_i - \sum_h \hat{w}^+_{ih})(s^+_j - \sum_h \hat{w}^+_{jh})$, noting that $\hat{e}^+_{ij} = e^+_{ij}$ when all $\hat{w}^+_{ij} = 0$. We then choose a random $a^+_{ij} = 1$, set $\hat{w}^+_{ij}$ to $w^+_{ij}$ of the same rank as the chosen $a^+_{ij}$, remove the chosen $a^+_{ij}$ and $w^+_{ij}$ pair from further consideration, and re-rank all remaining $w^+_{ij}$ and $a^+_{ij}$. We repeat this process until each positively weighted connections in the new network is associated with one of the original positive weights. Re-ranking at every step is important and allows convergence to the original strengths. In larger networks however, less frequent re-ranking may be performed without detriment to accuracy.

We now illustrate the effectiveness of our algorithm with an example. We consider networks of four equally sized modules with normally distributed weights centered at 0. Figure 3a,b shows a representative network, and an example null model of this network. Figure 3c shows the correlation coefficients between positive and negative strengths of the original network and its null model. These strength correlations exceed 0.9 for networks of size $\geq 100$, and exceed 0.95 for networks of size $\geq 250$ (solid lines). On the other hand, strength correlations effectively vanish when weights are randomly associated with connections, even if the degrees are preserved (broken lines). Figure 3d shows that the correlation coefficients between the positive and negative weights of the original and null model networks are very weak, which suggests an effective randomization. Strength correlations in empirical networks are likely to be even higher (see Results) because empirical networks typically have nonuniform degree distributions.

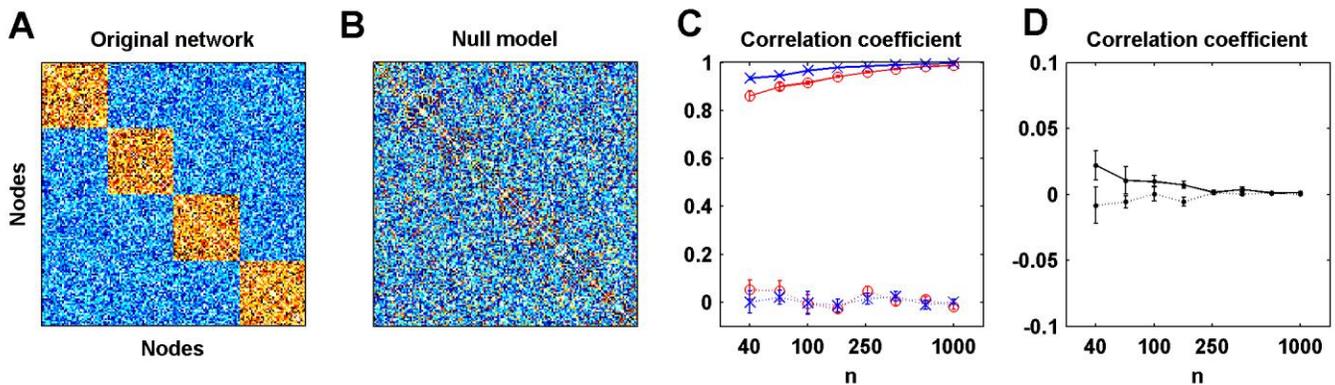

**Figure 3.** Degree-, weight- and strength-preserving null model. (a) A network with a maximal number of positive within-module connections (warm colors), a maximal number of negative between modules connections (cool colors), and a normally distributed weight distribution with mean of 0. (b) A null model of the network in (a), reordered by modular organization. (c) Correlation coefficients between positive (red) and negative (blue) strengths of the original and null model networks (solid lines), as a function of network size. Correlations between degree-preserving, but strength-non-preserving networks are shown for comparison (dashed lines). (d) Correlations between connection weights in the original and null-model networks. Error bars represent the standard error of the mean from 10 simulations.

## Results

In this section, we apply our measures to functional connectivity networks constructed from datasets in the 1000 Functional Connectomes Project (Biswal et al., 2010). The 1000 Functional Connectomes Project is an international open-access repository of resting-state functional connectivity MRI datasets. For the following analyses, we characterized functional connectivity networks from 18 available sites in this repository (Figure 4). The functional network associated with each site represents the group-average network of all subjects in that site. Each network contains 112 cortical and subcortical regions defined according to the Harvard-Oxford atlas, a probabilistic anatomical landmark-based atlas (Makris et al., 1999; Table S1). Demographic, recording and preprocessing details associated with these networks are described in Biswal et al. (2010) and references therein. The networks were kindly provided by Xi-Nian Zuo and Mike Milham, and are available on the project website (http://fcon_1000.projects.nitrc.org/).

### Goodness of high-modularity partitions

Network weights were near-normally distributed around a mean of 0 (Figure S1), and consequently all networks had large numbers of negative weights. We first considered the effect of these weights by studying properties of high- $Q^+$, $Q^*$, $Q^{GJA}$ and $Q^-$ partitions. The respective modularity measures incorporate differing influences of negative weights, from absent influence in $Q^+$ to sole influence in $Q^-$. Figure 5 shows properties of these high-modularity partitions averaged over all degenerate partitions of each network. Increasing influence of negative weights was associated with reduced numbers of modules and increased module size (Figure 5a,b), with larger numbers of within-module positive weights and a lower density of these weights (Figure 5c,d) and with increased non-rescaled contribution of negative weights (i.e. $Q^-$) and reduced non-rescaled contributions of positive weights (i.e. $Q^+$) (Figure 5e). High-$Q^-$ partitions were associated with very low values of non-rescaled contributions of positive weights (Figure 5e), and with very high numbers of discovered degenerate partitions (Figure 5f). These partitions were objectively less optimal than the other partitions, supporting our argument for a less important role of negative weights in partition determination.

Most current studies characterize functional network topology indirectly, by characterizing sparse and binary representations of these networks (Rubinov and Sporns, 2010). We next studied properties of high-modularity partitions of these representations, by considering binary networks with preserved 10%, 20%, 30%, 40% and 50% of the strongest positive connections. Figure 6 shows properties of high-modularity partitions associated with these networks, averaged over all degenerate partitions of each network. For meaningful comparison, we computed properties of these partitions in the original fully connected network; this allowed us, for instance, to examine organization of negative weights in these partitions even if the binary representations did not have negative weights. Figure 6 shows that reduced numbers of preserved connections in the binary networks was associated with increased numbers of modules and reduced module size (Figure 6a,b), with smaller numbers of within-module positive weights and a higher density of these weights (Figures 6c,d). Non-rescaled contributions of positive weights peaked when binary representations preserved 20% of strongest connections, while representations with fewer connections had low non-rescaled contributions of both positive and negative weights (Figure 6e) and high numbers of degenerate partitions (Figure 6f).

## Connectivity matrices

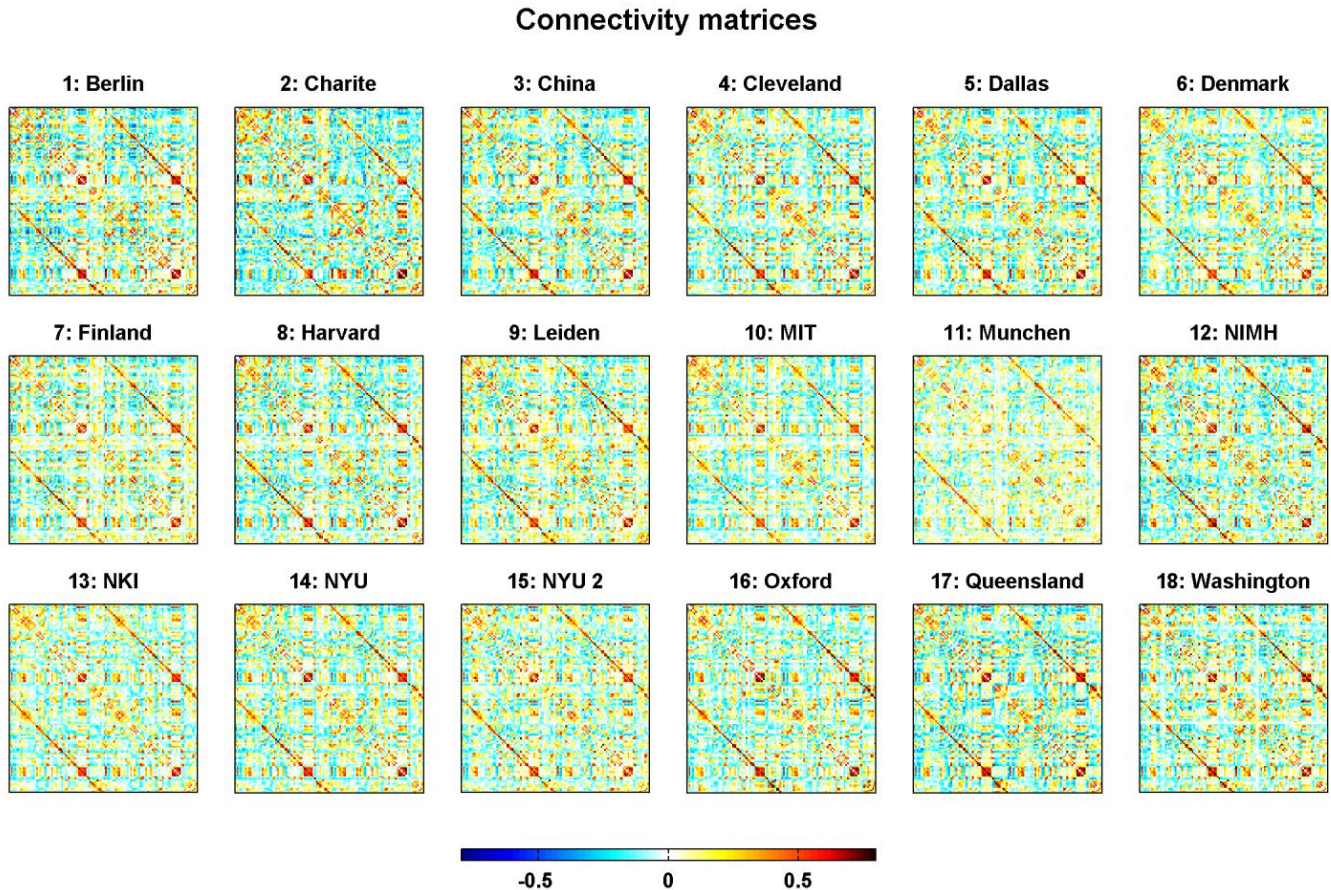

**Figure 4.** The examined resting-state functional MRI networks from the 1000 Functional Connectomes Project. All networks represent group averages over all subjects at each site, parcellated using the Harvard-Oxford Atlas. Node order is in Table S1.

We finally compared properties of observed high-modularity partitions to properties of high-modularity partitions of constructed null models. Null models had highly conserved strengths: the strength-strength correlation coefficient for empirical networks and their null model was $0.993 \pm 0.002$ for positive strengths and $0.996 \pm 0.001$ for negative strengths (mean $\pm$ standard deviation). Figures 5 and 6 show that null models had substantially smaller numbers of within-module positive weights, a lower density of these weights and lower non-rescaled contributions of positive and negative weights. These findings suggest that observed properties of high-modularity partitions of empirical networks were not associated with weight, degree and strength properties of these networks.

### Degeneracy of high-modularity partitions

Figures 5f and 6f show that the empirical networks had many degenerate high-modularity partitions. The number of discovered partitions varied for modularity measures and network topologies. Measures or topologies that ignored important weight information (by maximizing $Q^-$ or by discarding a large number of weights) had substantially larger numbers of degenerate partitions. On the other hand, null models of all networks had very few degenerate partitions. The number of degenerate partitions associated with weight-conserving characterizations may hence reflect inherently present degeneracy, but not spurious degeneracy due to loss of information.

Figures 5g and 6g show average distances between degenerate high-modularity partitions of the same network. In general, distances between partitions of empirical networks were much smaller than distances between partitions of corresponding null models. Despite this, distances between partitions of empirical networks remained substantial. For example, Figure 7a shows topologies of four representative high-$Q^*$ degenerate partitions. These partitions all have very similar $Q^*$ but differ in their organization, as reflected graphically and quantified with inter-partition distances. Figure 7b shows the distance matrix of all discovered high-$Q^*$ degenerate partitions of all empirical networks. Clusters on the diagonal represent distances between partitions of the same network, while clusters off the diagonal represent distances between partitions of different networks. Between-network partition distances were higher than within-network partition distances. Movie S1 loops through all discovered high-$Q^*$ degenerate partitions in

Figure 7b.

Figure 8 shows averages of all degenerate partitions of individual networks, and hence the within-module connection likelihoods in these networks. Degenerate partitions revealed a substantial module overlap in many networks. This overlap implies that certain regions or connections are not clearly associated with single modules.

Figure 9a shows the mean within-module connection likelihood matrix averaged over all matrices in Figure 8 and illustrates that many within-module connections were not consistently present between networks. Figure 9b shows the topology of connections that were present inside modules in more than 90% of cases.

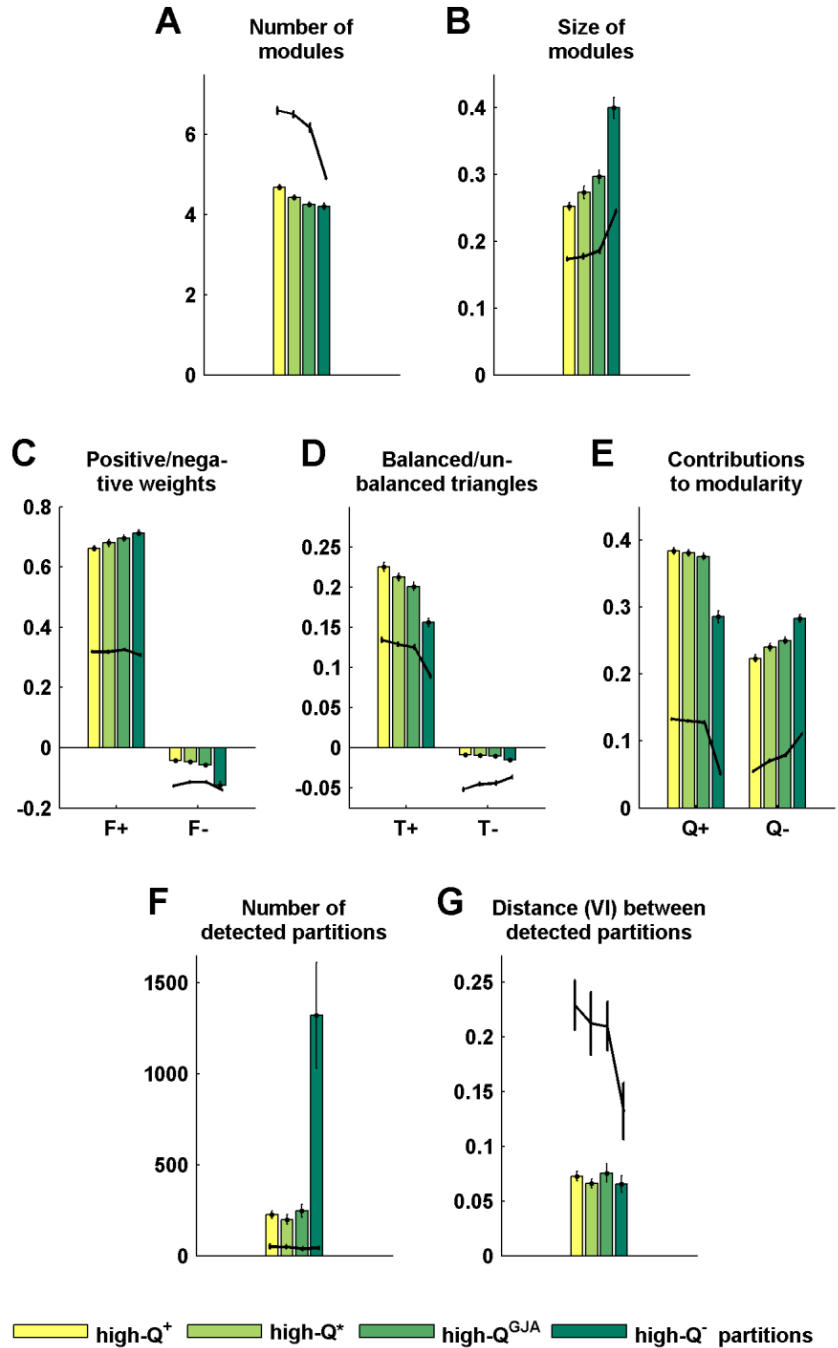

**Figure 5.** Properties of (from left to right) high-$Q^+$, $Q^*$, $Q^{GJA}$ and $Q^-$ partitions averaged over all degenerate partitions of all 18 networks: (a) number of modules, (b) size of modules (total number of within-module connections) as a proportion of network size, (c) number of positive and negative within-module weights, as a proportion of total within-module weights, (d) geometric mean of balanced and unbalanced within-module triangle weights, (e) non-rescaled contribution of positive and negative weights, (f) number of degenerate partitions, (g) distance between degenerate partitions of the same network. Black lines represent corresponding values in the null models. Error bars represent the standard error of the mean from 18 networks.

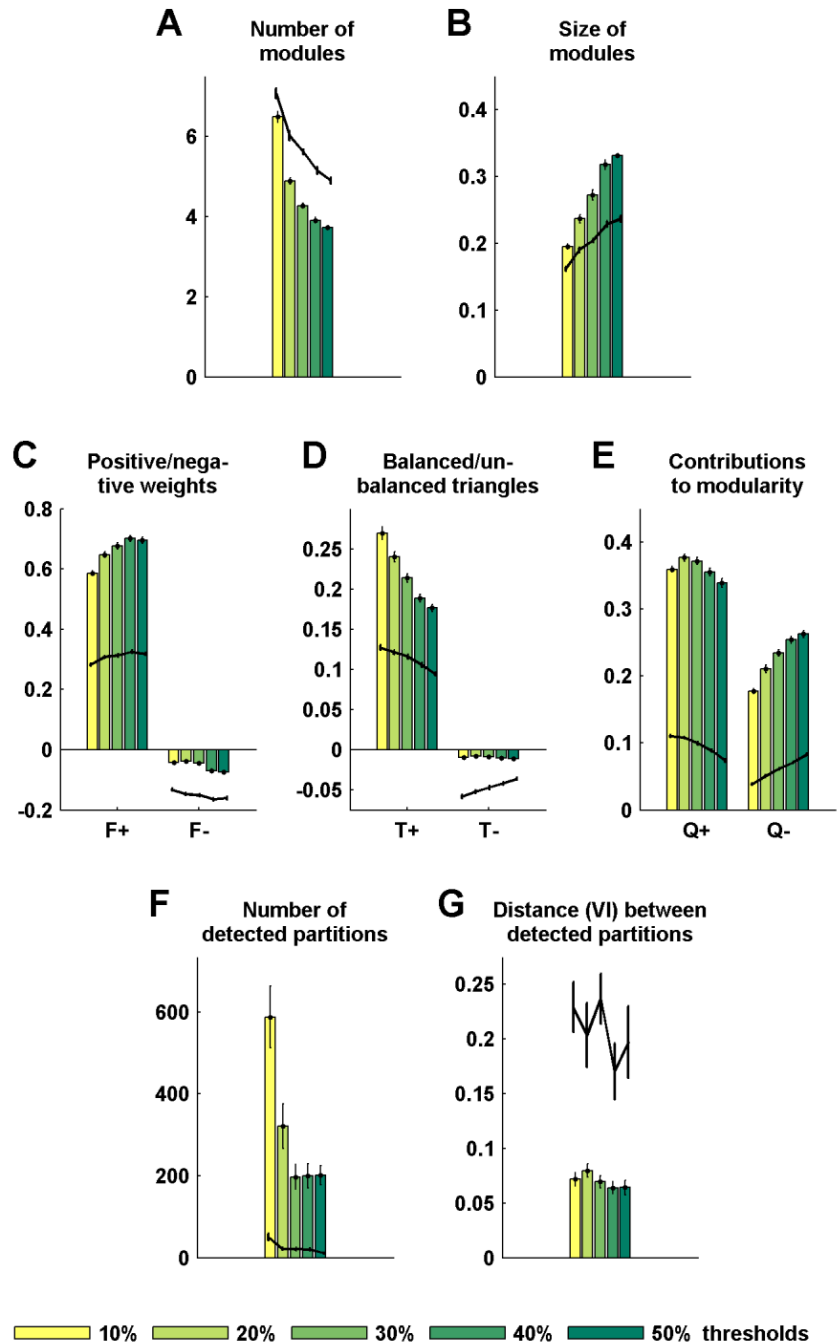

**Figure 6.** Properties of high-$Q^+$ partitions computed on binary networks with (from left to right) 10%, 20%, 30%, 40% and 50% of preserved connections. Panel descriptions are the same as in Figure 5.

These connections form four bilaterally symmetric modules (Table 1): 1) a large frontotemporoparietal module which includes primary somatosensory, motor and auditory areas; 2) a default-network module which includes parts of the prefrontal cortex, cingulate cortex and the inferior parietal lobule; 3) a temporolimbic module which includes visual association areas, amygdala and hippocampus; 4) a predominantly occipital module which includes the primary visual area. The discovered modules broadly agree with recent similar reports (Fair et al., 2009; He et al., 2009; Meunier et al., 2009) and our estimate of the default network agrees with convergent evidence for the composition of this network from multiple functional MRI approaches (Buckner et al., 2008). Our estimates of module composition are conservative as we only consider modules with highly stable connections. Consequently, we did not associate 41 regions (37% of all examined regions) with specific modules.

*Properties of regional centrality*

We finally examined the correlation between regional centrality measures within networks. As above, we considered measures that use positive or negative weights alone, measures that treat positive and negative weights symmetrically ($s'^+ - s'^-$, $h^+ - h^-$) and our proposed measures that treat positive and negative weights asymmetrically ($s'^*$, $h^*$). Figure 10a shows the correlation between the strength and diversity measures within individual partitions for these four cases, averaged over all degenerate partitions. Diversity measures were computed on their respective partitions (e.g. $h^+$ was computed on high-$Q^+$ partitions, while $h^*$ was computed on high-$Q^*$ partitions), but the choice of partition types did not qualitative change our results. Correlations between strength and diversity reflect the consistency of two complementary measures of regional importance and are clearly highest for the $s'^*$, $h^*$ pair of measures. Figure 10b-d shows scatter plots of regional strength and diversity, with data pooled from all 18 networks. Regions with simultaneously high $s'^*$ and $h^*$ were predominantly heteromodal, limbic and paralimbic (i.e. transmodal) in origin, and included the insula, the superior temporal cortex, the orbitofrontal cortex, the amygdala, the anterior cingulate cortex and the temporal pole.

## Discussion

We defined measures of modularity, high-modularity degeneracy and centrality in complex functional brain networks, and we described a null model against which these measures may be assessed. We applied our measures to an ensemble of resting-state functional MRI networks and demonstrated degeneracy of high-modularity partitions and strong correlations between two complementary measures of centrality.

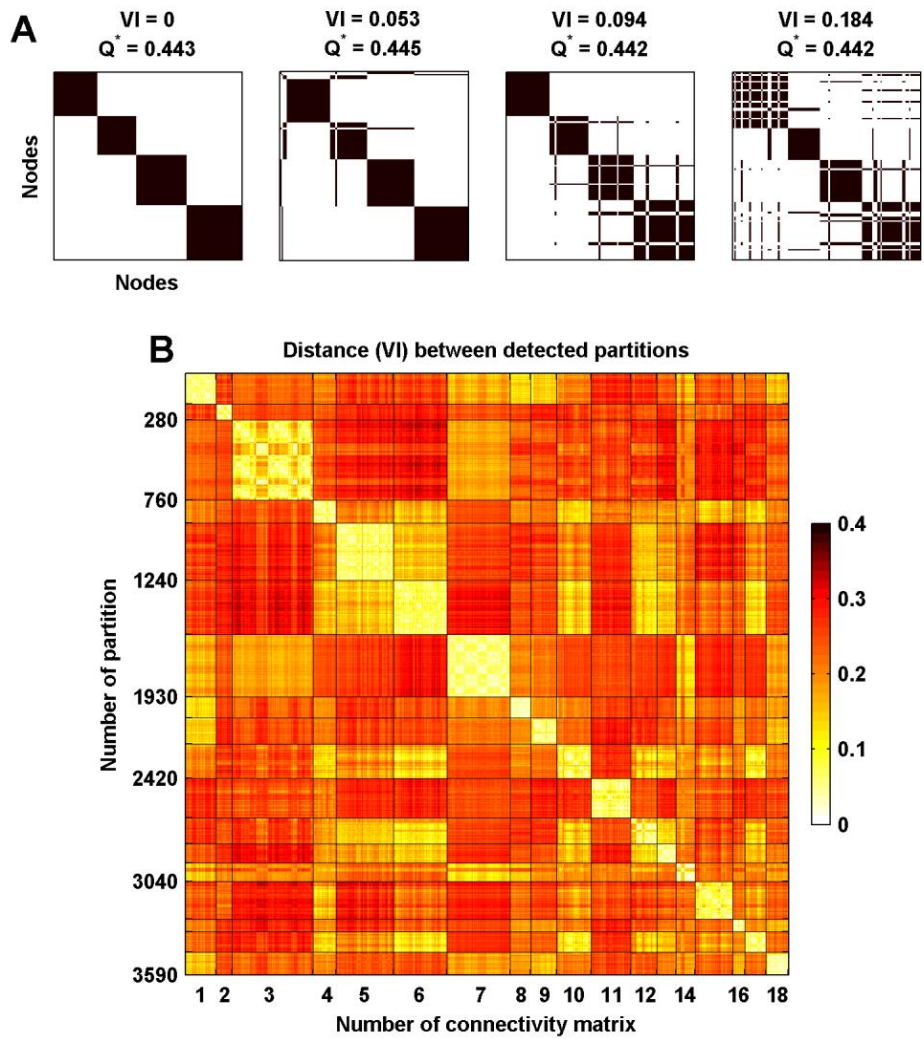

**Figure 7.** Exploration of high-$Q^*$ degenerate partitions. (a) Illustration of four degenerate partitions from the NIMH network. For each partition, labels specify the value of $Q^*$, and the distance to the left-most partition. (b) Distances between all discovered degenerate partitions of all networks.

## Within-module connection likelihood matrices

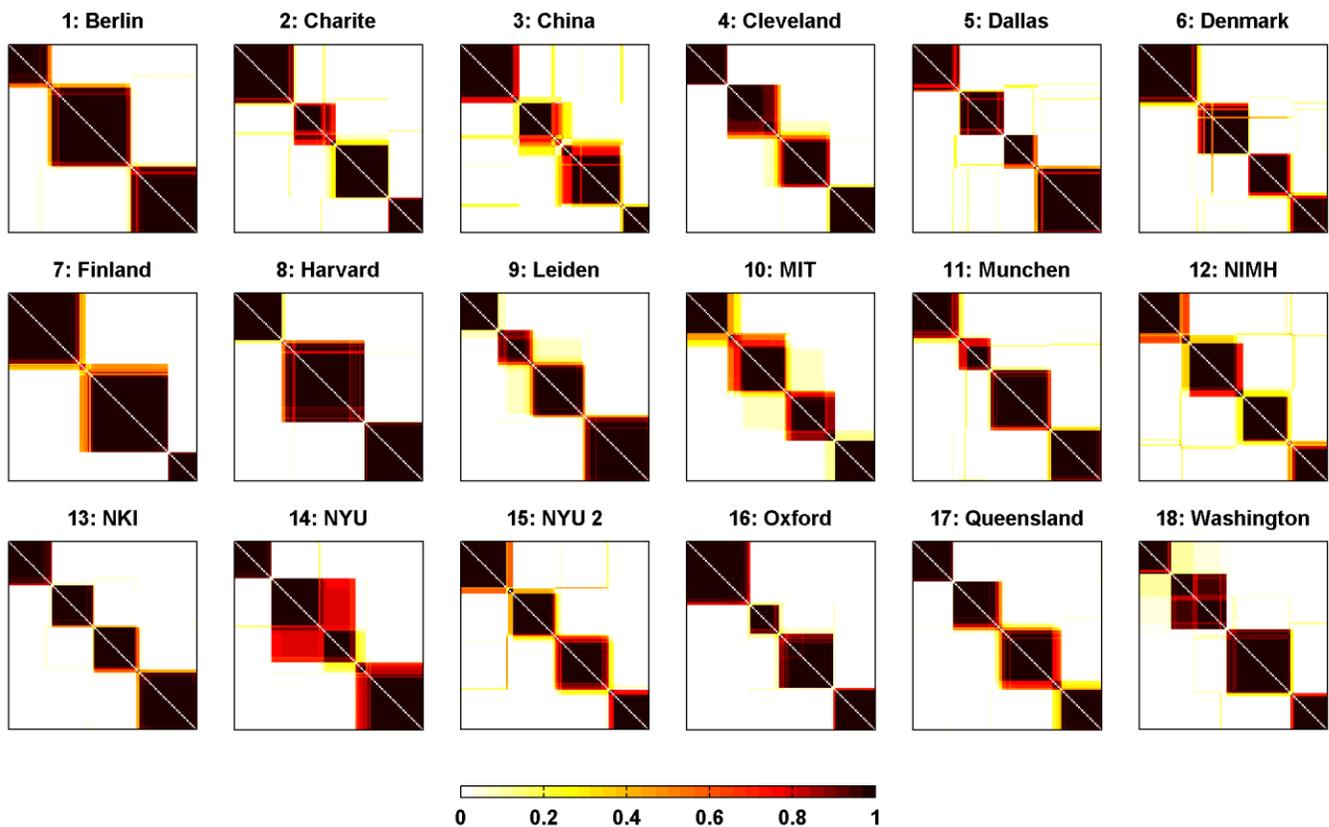

**Figure 8.** Within-module connection likelihood matrices, computed by averaging degenerate partitions within each network.

Our measures characterize fully connected, positively and negatively weighted functional networks. In contrast, most measures in current use characterize sparse and binary representations of functional networks. Sparse representations are typically constructed by defining arbitrary weight thresholds, require the exploration of many thresholds, and are consequently associated with multiple statistical comparison problems. Furthermore, thresholds are sometimes chosen not to preserve strong connections and remove weak connections, but to achieve levels of sparseness at which between-group differences are pronounced. In this study, we showed that partitions computed on increasingly sparse representations became increasingly less optimal. Our proposed measures may obviate the need for these arbitrary analyses and open the way towards more sound and reliable network characterizations.

Our measures incorporate negative weights but recognize the fundamentally different role of positive and negative weights in network organization. The nature of negative correlations in resting-state functional MRI remains controversial (Murphy et al., 2009; Fox et al.,

2009) although recent studies unambiguously demonstrate the neurophysiological origin of strong negative correlations (e.g. Chang and Glover, 2009). We note that our measures primarily depend on the relative difference between weight magnitudes and secondarily on the sign of the weights. For instance, when we linearly mapped the weight range $[-1,1]$ to the range $[0,1]$, such that all negative correlations were transformed into weak positive correlations, we found that properties of the resulting high-modularity partitions remained broadly similar (result not shown).

To the best of our knowledge, we are the first to report degeneracy of high-modularity partitions in functional brain networks. Analytical evidence for the presence of this degeneracy in real-world networks (Good et al., 2010) strongly suggests that degeneracy is a true feature of functional brain networks and not an artifact of noisy measurements or imprecise node definition. Current characterizations of the modularity in functional networks typically focus on only one high-modularity partition or report highly reproducible partitions (Joyce et al., 2010).

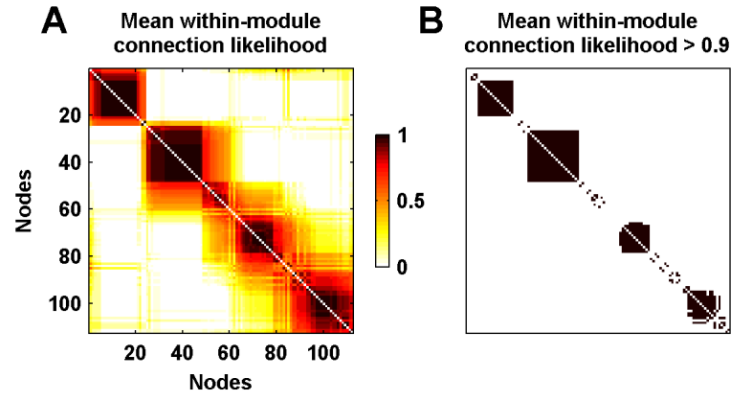

**Figure 9.** (a) Averaged within-module connection likelihood matrices from Figure 8. (b) Binary representation of the matrix in (a), showing connections with greater than 90% within-module likelihood. Table 1 lists regions in the four large connected components of this matrix.

TABLE 1: Consistent modules in 18 examined networks.

| Region | Left | Right |
|---|---|---|
| *Module 1* | | |
| Precentral Gyrus | X | X |
| Supplementary Motor Cortex | X | X |
| Postcentral Gyrus | X | X |
| Supramarginal Gyrus, anterior | X | X |
| Superior Temporal Gyrus, anterior | X | |
| Superior Temporal Gyrus, posterior | X | X |
| Heschl's Gyri | X | X |
| Planum Polare | X | X |
| Planum Temporale | X | X |
| Insular Cortex | X | X |
| Central Opercular Cortex | X | X |
| Parietal Opercular Cortex | X | X |
| *Module 2* | | |
| Frontal Pole | X | X |
| Superior Frontal Gyrus | X | X |
| Middle Frontal Gyrus | X | X |
| Inferior Frontal Gyrus, pars triangularis | X | |
| Paracingulate Gyrus | X | X |
| Caudate | X | X |
| Angular Gyrus | X | X |
| *Module 3* | | |
| Temporal Pole | X | X |
| Temporal Fusiform Cortex, anterior | X | X |
| Temporal Fusiform Cortex, posterior | X | X |
| Middle Temporal Gyrus, anterior | X | X |
| Inferior Temporal Gyrus, anterior | X | X |
| Amygdala | X | X |
| Hippocampus | X | X |
| Parahippocampal Gyrus, anterior | X | X |
| Parahippocampal Gyrus, posterior | X | X |
| *Module 4* | | |
| Lingual Gyrus | X | X |
| Cuneal Cortex | X | X |
| Lateral Occipital Cortex, inferior | X | X |
| Supracalcarine Cortex | | X |
| Intracalcarine Cortex | X | X |
| Temporal Occipital Fusiform Cortex | X | X |
| Occipital Fusiform Gyrus | X | X |
| Occipital Pole | X | X |

Modules were bound by within-module connections present in more than 90% of examined partitions (see Figure 9).

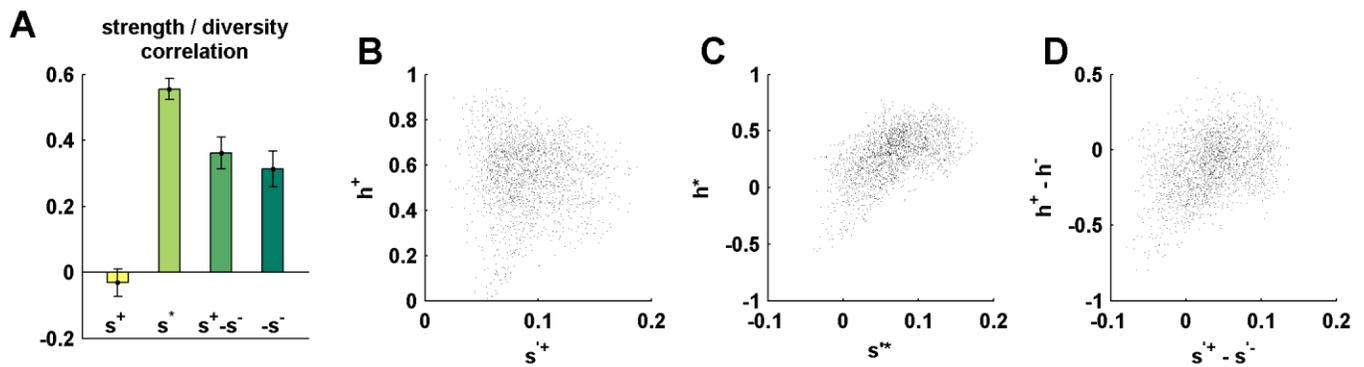

**Figure 10.** (a) Correlation between (from left to right) $s'^+$ and $h^+$, $s'^+$ and $h^+$, $(s'^+ - s'^-)$ and $(h^+ - h^-)$, $-s'^-$ and $-h'^-$, averaged over all discovered degenerate partitions. Error bars represent the standard error of the mean from 18 networks. (b)-(d) Scatter plots of strength and diversity values, pooled from all networks (the $-s'^-$ and $-h'^-$ plot is not shown).

We argue that such characterizations are incomplete for two reasons. Firstly, a single observed modularity partition is unlikely to represent a true global optimum because modularity maximization algorithms are approximate and because observed partitions are subject to recording and preprocessing distortions. Secondly and more importantly, degenerate partitions have the potential to collectively capture dynamic regional interactions that cannot be possibly detected with single partitions. These interactions may include context-dependent regional activation in distinct functional systems or switching between distinct systems in the resting state (e.g. Bressler and McIntosh, 2007). Our consideration of degenerate partitions led us to consider the likelihood with which connections bind nodes into coherent modules. This view is related to notions of cluster stability (Bellec et al., 2010), connection-based modules (Ahn et al., 2010) and overlapping modules (Palla et al., 2005).

We purposefully restricted network characterization to measures with simple neurobiological interpretations. It is important to emphasize that connections in cross-correlation-based functional networks do not occupy physical space and typically represent the dynamic outcome of numerous direct and indirect network interactions (Koch et al., 2002; Honey et al., 2009). Several popular measures of network topology are difficult to interpret in these networks. For instance, measures of wiring cost are difficult to interpret because sparse anatomical topologies can generate highly coherent and "costly" functional patterns (Vicente et al., 2008). Measures based on paths, such as characteristic path length and small-worldness, are also difficult to interpret because functional networks are fully connected and statistical relationships between all pairs of regions are already directly expressed by connection weights between these regions. Short characteristic path lengths may reflect a large number of inter-modular connections, but not necessarily correspond to high functional integration, as is often suggested. Such ambiguities underscore the desirability of combining functional connectivity with an underlying structural connectivity model (Sporns et al., 2005) and are likely to become more apparent with further characterization of increasingly high-resolved and fully-connected functional networks. We hope that our proposed approach will enable more meaningful characterizations of functional networks across conditions or subjects and will stimulate further discussion on the use and interpretation of functional brain network measures.

### Acknowledgments

We are very grateful to Mike Milham and Xi-Nian Zuo for making the resting-state functional connectivity datasets used in this study freely available, and to Ben Good for providing detailed feedback on an earlier version of the manuscript. We thank Dani Bassett, Jonathan Power and James Roberts for stimulating discussions. The study was funded in part by the Human Connectome Project (1U54MH091657-01) from the 16 NIH Institutes and Centers that support the NIH Blueprint for Neuroscience Research. MR and OS were supported by the JS McDonnell Foundation. MR was supported by the CSIRO ICT centre.